# Mass distribution of magnetized quark-nugget dark matter and comparison with requirements and observations

J. Pace VanDevender[1]*, Ian M. Shoemaker[2], T. Sloan[3], Aaron P. VanDevender[1], Benjamin A. Ulmen[4]

[1]VanDevender Enterprises LLC, 7604 Lamplighter LN NE, Albuquerque, NM 87109 USA. [2]Department of Physics, Virginia Tech, Blacksburg, VA 24061, USA. [3]Department of Physics, Lancaster University, Lancaster, LA1 4YB, UK. [4]PO Box 5800 MS-1159, Sandia National Laboratories, Albuquerque, NM 87185-1159

*pace@vandevender.com

**Abstract**

Quark nuggets are a candidate for dark matter consistent with the Standard Model. Previous models of quark nuggets have investigated properties arising from their being composed of strange, up, and down quarks and have not included any effects caused by their self-magnetic field. However, Tatsumi found that the core of a magnetar star may be a quark nugget in a ferromagnetic state with core magnetic field $B_{surface} = 10^{12\pm1}$ T. We apply Tatsumi's result to quark-nugget dark-matter and report results on aggregation of magnetized quark nuggets (MQNs) after formation from the quark-gluon plasma until expansion of the universe freezes out the mass distribution to ~$10^{-24}$ kg to ~$10^{14}$ kg. Aggregation overcomes weak-interaction decay. Computed mass distributions show MQNs are consistent with requirements for dark matter and indicate that geologic detectors (craters in peat bogs) and space-based detectors (satellites measuring radio-frequency emissions after passage through normal matter) should be able to detect MQN dark matter. Null and positive observations narrow the range of a key parameter $B_o \sim B_{surface}$ to $1 \times 10^{11}$ T $< B_o \leq 3 \times 10^{12}$ T.

**Introduction**

Eighty-five percent of the mass in the universe is dark matter [1-5]. Its nature is unknown. Extensive searches for an explanation beyond the Standard Model [6] have not detected any statistically significant signals [4]. Macroscopic quark nuggets are a candidate for dark matter composed approximately equal numbers of strange, up, and down quarks and are consistent with the Standard Model. They are called many names [7-12]. All quark nuggets interact [13-15] through gravitational and strong nuclear forces. A brief summary of quark-nugget formation, stability, and compliance with dark-matter requirements [7-35] has been updated from Ref. 17 and is provided for completeness as Supplementary Note: Quark-nugget research summary.

Since quark nuggets are composed of three quarks, they are baryons. Convention often assumes that all baryons are normal (i.e. not dark) matter and designates the number density and mass density of normal matter as $n_b$ and $\rho_b$ respectively. However, since quark nuggets are baryons too, we use $n_n$ and $\rho_n$ for those quantities of normal matter, respectively.

Most previous models of quark nuggets have not considered effects caused by their self-magnetic field. However, Tatsumi [16] explored the internal state of quark-nugget cores in magnetars and found they may exist as a ferromagnetic liquid with a surface magnetic field $B_{surface} = 10^{12\pm1}$ T. His theory uses the bag model [21] because more rigorous lattice and perturbative calculations with chromodynamics are intractable for the relevant energy scale of ~ 90 MeV. Tatsumi find that a stable ferromagnetic state should exist if the coupling constant $\alpha_c$ is ~ 4 at this energy scale. A metastable ferromagnetic state may exist for $\alpha_c > 1.75$. Although $\alpha_c$ gets larger as the energy scale decreases, it is not known if $\alpha_c$ is indeed that large. Even with those considerations, his conclusions have important consequences and are testable through searches for quark-nugget dark matter. We apply his ferromagnetic-liquid model to quark-nugget dark matter. In a previous paper, we found the self-magnetic field strongly enhances the interaction cross section of a magnetized quark nugget (MQN) with a surrounding plasma [17]. In this paper, we explore the interaction between MQNs, which are magnetically attracted into aggregating collisions. As ferromagnetic liquids, they will combine and remain strongly magnetized after aggregating.

Mass distributions have been computed and are reported in this paper as a function of the key parameter $B_o$ that is related to the average value $<B_{surface}>$:

$$<B_{surface}> = \left(\frac{\rho_{QN}}{10^{18}(kg/m^3)}\right)\left(\frac{\rho_{DM\_T=100Mev}}{1.6x10^8(kg/m^3)}\right)B_o. \quad (1)$$

Throughout this paper, we will use $B_o$ as a key parameter. The value of $B_o$ equals $<B_{surface}>$ if the mass density of MQNs $\rho_{QN} = 10^{18}$ kg/m$^3$ but that number is quite uncertain. Witten's [7] estimate of "somewhat greater than nuclear density" and a rough calculation that gives ~ $7.5 \times 10^{17}$ kg/m$^3$ is consistent with $6 \times 10^{17}$ to $7 \times 10^{17}$ kg/m$^3$ covering the range of uncertainty in the proton radius and the corresponding mass density. Peng, *et al.*'s [30] more recent work covers a range of $1.7 \times 10^{17}$ to $3.3 \times 10^{18}$ kg/m$^3$ for quark matter in quark stars. We use $\rho_{QN} = 1 \times 10^{18}$ kg/m$^3$ in the calculations below. In addition, the $B_o$ parameter depends on the density of dark matter $\rho_{DM} = 1.6 \times 10^8$ kg/m$^3$ at time t ≈ 65 μs, when the temperature T ≈ 100 MeV in accord [36] with the standard ΛCDM cosmology. If more accurate values of $\rho_{DM}$ when T ~ 100 MeV, $\rho_{QN}$, or $B_o$ are found, then equation (1) can give a correspondingly more accurate value of $<B_{surface}>$.

We assume MQNs are formed with baryon number $A = 1$ at the beginning of baryogenesis, as are protons and neutrons, and aggregate by binary collisions, similar to nucleogenesis of low-A elements in the standard cosmological model. Since quark-nuggets are electrically neutral (or neutralized as discussed below) and quickly aggregate to $A \gg 1$ (as shown below), we assume they are decoupled from the thermal environment of the co-moving universe. With those assumptions, we simulated the aggregation of MQNs from the time baryons form in the quark-gluon plasma until expansion of the universe reduces MQN density and freezes out the mass distribution.

Although Tatsumi's theory of ferromagnetism is applicable to quark-nuggets with $A \gg 1$, we note that the magnetic moments and mass densities of neutrons and protons, which are also baryons with $A = 1$, correspond to magnetic fields $B_o = 2.5 \times 10^{12}$ T and $1.5 \times 10^{12}$ T respectively. These fields are in the middle of the range identified by Tatsumi, so the aggregation from $A = 1$ is not unreasonable.

The results are strongly dependent on surface magnetic field $B_{surface}$ from Tatsumi's theory. We find the strong magnetic interaction of MQNs and their resulting mass distribution are consistent with requirements for dark-matter candidates, as summarized by Jacobs, Starkman, and Lynn [12]: Large scale structure constraints (cosmic microwave background and the Bullet Galaxy) from self-interactions and interactions with other baryons and photons, ancient mica observations [32], and results from the Skylab experiment [36]. Additional requirements may be added in the future. We conclude that MQNs provide a candidate for dark matter that appears to be consistent with the Standard Model of particle physics, without any extensions and merit further investigations.

Null results from a systematic search for MQNs with a 30 sq-km section of the Great Salt Lake that was monitored with an array of three hydrophones are presented in the Discussion. Results [37] from investigation of episodic non-meteorite impacts are consistent with an MQN impact and are also briefly summarized in the Discussion section. The null and episodic results consistent with the mass distributions for non-excluded values of $B_o$ motivate additional, systematic investigations [38].

We use standard international MKS units except temperature, which is expressed in electron volts (eV), and interaction strengths $\sigma_x/m_x$ (i.e. cross sections divided by mass) which is expressed in cm$^2$/g to facilitate comparison with values quoted in the literature.

**Results**

**Direct Simulation Monte Carlo computation of MQN formation and aggregation**

The mass distribution of magnetized quark nuggets, existing as a ferromagnetic fluid held together with the strong force as described by Tatsumi [16], is computed from their aggregation in binary collisions under the influence of their self-magnetic fields. We assume that singlets with $A = 1$ are formed when the thermal energy of the early universe is much less than the rest mass of the singlet ($kT_f \ll m_s c^2$) for Boltzmann constant k, formation temperature $T_f$, singlet mass $m_s$, and speed of light c. Protons and neutrons are similarly formed in the $\Lambda$CDM model [1]. Since $m_s$ is somewhat larger [7, 8, 11] than the proton mass, we set $m_s = 1.7 \times 10^{-27}$ kg and choose $T_f = 100$ MeV, which corresponds to time t ~ 65 µs and is compatible with the formation of particles of mass $m_s$ from the quark-gluon plasma [39].

MQNs interact with each other through their magnetic fields and aggregate through binary collisions when their magnetic potential energy is greater than their initial kinetic energy. Because the magnetic dipole-dipole interaction scales as $r^{-3}$, there is no centrifugal force barrier to a direct collision and aggregation. Since MQNs are a ferromagnetic liquid in Tatsumi's theory [16], the domains align after aggregation and the surface magnetic field is preserved, so $B_o$ is preserved in aggregations.

MQNs reorient as they approach each other to experience maximum attractive force, which is the orientation with minimum potential energy [40]. Therefore, our simulation does not track orientation of each MQN.

Direct Simulation Monte Carlo (DSMC) methodology was developed for the aggregation of particles in a collisionless fluid. We adapted the procedure described by Kruis, Maisels, and Fissan [41] who verified the methodology by comparing their results with analytic solutions of particle aggregation. Our adaptation is presented in the Methods section.

With both net electric and magnetic fields, the aggregation cross section $\sigma_{EM} = \pi\, r_{EM}^2$ for the collision of particles $i$ and $j$, and $r_{EM}$ is given implicitly by the sum of the electric and magnetic potentials exceeding the kinetic energy at infinity:

$$\left(\frac{1}{4\pi\varepsilon_0}\right)\frac{q_i q_j}{r_{EM}} - \frac{\mu_o m_{m,i} m_{m,j}}{4\pi r_{EM}^3} > \frac{1}{2}\frac{m_i m_j |\vec{u}_i - \vec{u}_j|^2}{m_i + m_j} \qquad (2)$$

in which $q_i$ and $q_j$, $m_{m,i}$ and $m_{m,j}$, $m_i$ and $m_j$, and $\vec{u}_i$ and $\vec{u}_j$, are, respectively, the electric charges, magnetic dipole moments, masses, and vector velocities of the i and j particles. As usual, $\varepsilon_o$ is vacuum permittivity and $\mu_o$ is vacuum permeability. The middle term is the relative magnetic potential for two particles with different magnetic dipole moments separated by the distance $r_{EM}$ [41].

Since Tatsumi finds this ferromagnetic configuration is a liquid held together by the strong nuclear force, MQNs will be highly spherical. Consequently, we approximate the magnetic moment $m_m$ of each spherical, uniformly magnetized quark nugget as the magnetic moment of a current loop with the same radius $r_{QN}$ and magnetic field $B_o$ as the quark nugget, at a distance $r_{QN}$ on axis and above the center of the loop. We assume each quark nugget is a sphere with uniform mass density $\rho_{QN}$, so the mass $m_{QN}$ of each quark nugget is $(4/3)\pi\, \rho_{QN}\, r_{QN}^3$ and

$$m_m = \frac{3 B_o m_{QN}}{\mu_o \rho_{QN}}. \qquad (3)$$

Substituting equation (3) into equation (2), multiplying each side by $r_{EM}^3$, and rearranging gives a cubic equation in $r_{EM}$. This is then calculated for each collision using the usual formula for the root of a cubic equation.

We assume that each singlet is formed with $A = 1$, so each contains one up, one down, and one strange quark. The assembly is in the theoretically predicted, ultra-dense, color-flavor-locked (CFL) phase [18] of quark matter. Steiner, *et al.* [19] showed that the ground state of the CFL phase is color neutral and that color neutrality forces electric charge neutrality. Models of unmagnetized quark nuggets by Xia, *et al.*[11] and by Zhitnitsky, *et al.* [42] predict different small (i.e. the ratio of electric charge per baryon mass is much less than one), non-zero internal electric charges, but both predict charge neutralizing surface layers. In addition, the magnetic field of MQNs significantly changes their internal energy [29, 43]. Including the magnetic field in the equilibrium calculation [44] reduces the internal electron per baryon ratio to ~0.0003 for nuclear density quark nuggets and provides electric charge neutrality. Therefore, we assume the net internal charge plus the electrically neutralizing surface layers produce zero net electric charge.

The aggregation cross section is considerably simplified for zero net electric charge:

$$\sigma_{ij} = \left( \frac{9\pi^{1/2} B_o^2}{2\mu_o \rho_{QN}^2} \right)^{2/3} \left( \frac{m_i + m_j}{|\vec{u}_i - \vec{u}_j|^2} \right)^{2/3}. \tag{4}$$

The DSMC method for computing aggregations [41] uses the cross sections $\sigma_{ij}$ for every particle pair $(i, j)$ to simulate the aggregation process, as explained in more detail in the Methods section.

The simulation starts with 100,000 particles, as recommended by Kruis, Maisels, and Fissan [41] to provide adequate statistics. The initial speeds of the particles were generated to fit a Maxwell-Boltzmann distribution with temperature $T = 100$ MeV, and their velocity vectors were generated by generating velocity unit vectors with random orientation in space. Approximately 96% of the initial velocities have relativistic $\gamma < 2.0$ and 61% have $\gamma < 1.2$. Less than 0.1% have $\gamma \geq 5$.

Quark nuggets are decoupled from the temperature of the universe when they are formed. They interact only with other quark nuggets in binary collisions that conserve linear momentum. Net angular momentum from their collisions is radiated away by their rotating magnetic field, so their average energy per unit mass decreases as they aggregate.

Approximately $10^{-14}$ s into the simulation, which is the beginning of the 10$^{th}$ generation, aggregations have reduced the velocities to 99.6% with $\gamma < 2.0$ and 96.5% with $\gamma < 1.2$. Less than 0.02% have $\gamma \geq 5$. Even though some of the initial velocities are marginally relativistic, we use non-relativistic dynamics to do these first calculations to minimize computing time.

Selection and aggregation of $i$-particle, $j$-particle $(i, j)$ pairs proceed for 50,000 aggregations, which we call one generation. At the end of each generation, only 50,000 particles remain. Since mass is conserved, the mean mass has doubled. At the completion of each generation, the time within the simulation and all data are saved for analysis and for restart if needed. The volume of the simulation is then doubled by expanding the simulated volume to include the surrounding space. Since the aggregation process is uniform in space, the mass and velocity distributions of particles in that additional volume are quite similar to those of the 50,000 remaining particles. Consequently, we duplicate the mass and speed of each of the 50,000 particles to restore the number of particles to 100,000 and maintain adequate statistics. Since elastic scattering randomizes the directions of the velocity vectors, we randomize the directions of the velocities for the duplicated particles and, therefore, avoid dividing by zero as the next generation is simulated.

As described in the Methods section, the DSMC process that calculates each aggregation depends on
1. aggregating collision rates for each of $n$ particle pairs $(i, j)$, with mass $m_i$ and $m_j$, respectively,
2. velocities $\vec{u}_i$ and $\vec{u}_j$, respectively,
3. aggregation cross section $\sigma_{ij}$, which is given by equation (4),
4. co-moving volume V of the simulation,
5. dark-matter mass density $\rho_{DM}$ at the time of each aggregation, and
6. parameters $B_o$ and $\rho_{QN}$, which are independent of time.

The aggregation calculation is described in the Methods section and produces a mass and velocity distribution at the end of each generation. The process also calculates the time increment $\delta_{g,k}$ associated with each $k = 0$ to $40{,}999$ aggregation in each generation $g$. The time increment $\delta_g$ associated with the $g^{th}$ generation is the sum over $k$ of $\delta_{g,k}$ values in the $g^{th}$ generation.

$$\delta_g = \sum_{k=0}^{\frac{n}{2}-1} \delta_{g,k} = \sum_{k=0}^{\frac{n}{2}-1} \frac{2}{\rho_{DM\_g,k}} \left( \frac{2\mu_o \rho_{QN}^2}{9\pi^{1/2} B_o^2} \right)^{2/3} f_{g,k}(m_i, m_j, \vec{u}_i, \vec{u}_j) \ . \tag{5}$$

The function $f_{g,k}(m_i, m_j, \vec{u}_i, \vec{u}_j)$ is derived in the Methods section.

Since the co-moving volume $V$ = total dark-matter mass divided by $\rho_{DM}$ is continually changing with time, each aggregation $k$ depends on the time-varying dark-matter mass density $\rho_{DM}(t) = \rho_{DM\_g,k}$ in equation (5). Although $B_o$ is a constant for a simulation, its value changes $\delta_{g,k}$, which changes $\rho_{DM\_g,k}$. Therefore, a simulation should be completed for each $B_o$. Each simulation requires several months of computing.

However, we found that $\rho_{DM}$, is approximately constant within the 50,000 aggregations of a generation until the aggregations and expansion of the universe begin to freeze out the mass distribution. Once the aggregation rate slows, freeze-out quickly follows. The result is that assuming $\rho_{DM\_t}$ is constant within a generation is sufficient for specifying the mean and maximum mass in a distribution to within one generation, which is within a factor of 2. As shown in Fig. 5, mean mass and maximum mass are such a strong function of $B_o$, an uncertainty of a factor of 2 in mass gives an uncertainty of only 8% in the corresponding $B_o$. Eight percent uncertainty is sufficient to compare predictions with observations to narrow the range of $B_o$ from Tatsumi's $10^{12 \pm 1}$. We found that representing each mass distribution as a sum of the number of masses in each decade of mass is sufficient for this purpose. So the uncertainty in $B_o$ dominates the uncertainty in the mass distribution. Once the value of $B_o$ is determined to within +/- 20%, a full simulation should be run for representative values of $B_o$.

The assumption of constant $\rho_{DM\_g,k}$ within a generation $g$ allows approximate mass distributions to be calculated by post processing the results from one detailed simulation with $B_o = 10^{12}$ T. Equation (5) is approximated by

$$\delta_g \sim \frac{2}{\rho_{DM,g}} \left( \frac{2\mu_o \rho_{QN}^2}{9\pi^{1/2} B_o^2} \right)^{2/3} \sum_{k=0}^{\frac{n}{2}-1} f_{g,k}(m_i, m_j, \vec{u}_i, \vec{u}_j) \ . \tag{6}$$

The time $t_G$ since the beginning of the universe and associated with the completion of the $G^{th}$ generation equals the 65 μs start time plus the sum of the generation times $\delta_g$ for generations up to and including the $G^{th}$ generation.

$$t_G = 6.5 \times 10^{-6} + \sum_{g=1}^{G} \delta_g \ . \tag{7}$$

In post processing, $\rho_{DM}(t) = \rho_{DM\_g,k}$ is updated to be consistent with the standard ΛCDM Model, as explained in the next paragraph. The time $t_G$ is also associated with the mass and velocity distributions calculated in the $G^{th}$ generation. The process continues until $t_G \geq 13.8$ Gyr, the present age of the Universe.

We use a solution [45] of the Friedmann Equation [46] in a radiation-dominated Universe to obtain the temperature T as a function of time t. The solution includes the time dependent effective number of relativistic degrees of freedom which would have the same entropy density at the same photon temperature [39]. The solution is appropriate for time t up to ~ 50,000 years, which is well beyond the freeze-out time of the MQN mass distribution. Fitting that solution to a simple analytic function lets the simulation run as quickly as possible. The fit is adequate to +/- 5% in T for 65 μs ≤ t ≤ 122 ms, by which time the mass distributions have frozen out, and to +/- 13 % for 122 ms < t ≤ $10^{10}$ s. For example, for $B_o = 1.085 \times 10^{12}$ T, time goes from 3 μs generation 89 to 13.8 Gyr in generation 90; therefore, the error after t = 55,000 years is well within the factor of 2 uncertainty in mean mass and 8% uncertainty in $B_o$ described above.

We use the Standard Model and the current values of cosmological variables [1] (i.e. number density $n_\gamma$ of photons in thermal equilibrium, the ratio of normal-matter number density $n_n$ to photon number density $n_\gamma$, and the ratio of cold-dark-matter mass density $\rho_{DM}$ to normal-baryon mass density $m_p n_n$) to calculate the time dependent dark-matter density $\rho_{DM}(t)$ from temperature $T_{MeV}(t)$:

$$T_{MeV} = 1.042 t^{-0.4826}$$
$$n_\gamma = 16\pi\varsigma(3)\left[\frac{kT}{hc}\right]^3 = 3.15 \times 10^{37} T_{MeV}^3$$
$$n_n \simeq 6.15 \times 10^{-10} n_\gamma \quad (8)$$
$$\rho_{DM} \simeq 5.38 m_p n_n$$
$$\rho_{DM} = 167.0 T_{MeV}^3 = 188.9 t^{-1.448}$$

in which Rieman Zeta function $\varsigma(3) = 1.202$ and the temperature $T_{MeV}$ in MeV as a function of time t in seconds..

The DSMC technique made it possible to simulate the aggregation of quark nuggets from ~65 μs after the universe began to the present day in about three months of processing on a personal computer.

**Aggregation overcomes decay by weak interaction**

Since quark nuggets with baryon number $A = 1$ are not observed in accelerator experiments, they decay through the weak interaction in ~ 0.1 ns unless their decay is interrupted. We find that the aggregation time is typically 0.003 ns, so aggregation overcomes decay and allows growth of quark nuggets to $A \gg 1$, where they are magnetically stabilized.

**Aggregated mass distributions of electrically neutral or neutralized MQNs**

The aggregation process proceeds quickly, as shown in Fig. 1. By t = 1 ms, the mass distribution of electrically neutral or neutralized MQNs has essentially frozen out and changes less than factor of ~2 for the next 10 Gyr.

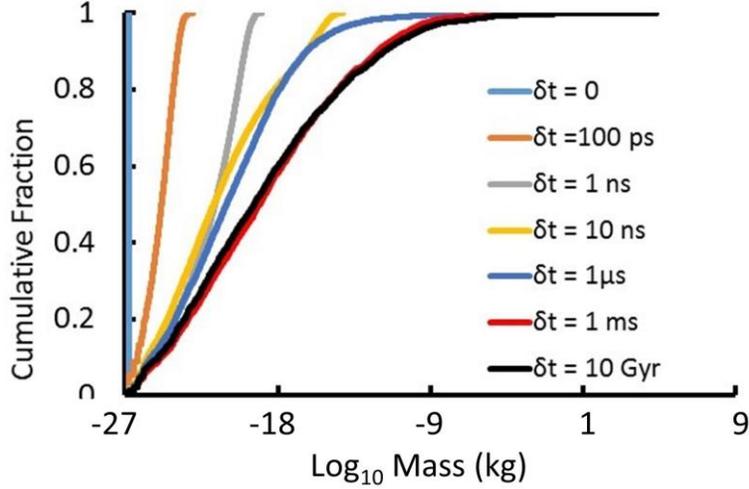

**Figure 1.** Snapshots of the masses at various times $\delta t$ after the beginning of the simulation at $t = 65$ μs and for $B_o = 10^{12}$ T. Distribution changes from a single mass = $1.6 \times 10^{-27}$ kg at $\delta t = 0$ to an increasingly broad mass distribution with increasing time.

The simulated mass distribution evolves quickly between the assumed formation of singlets with baryon number $A = 1$ at $t = 65$ μs, when temperature $T = $ ~100 MeV, and 1 ms, when the largest mass is 8 kg. By $t = 10$ Gyr, the largest mass is 20,800 kg and the rest of the distribution has changed little.

Computed quark-nugget mass distributions for electric-charge-neutral collisions and the time-dependent dark-matter density $\rho_{DM}(t)$ given by equation (8) are shown in Fig. 2 as the cumulative fraction $F_M$ of the mass distribution between 0 and mass $M$:

$$F_M = \int_0^M \frac{dn_m}{dm} dm \tag{9}$$

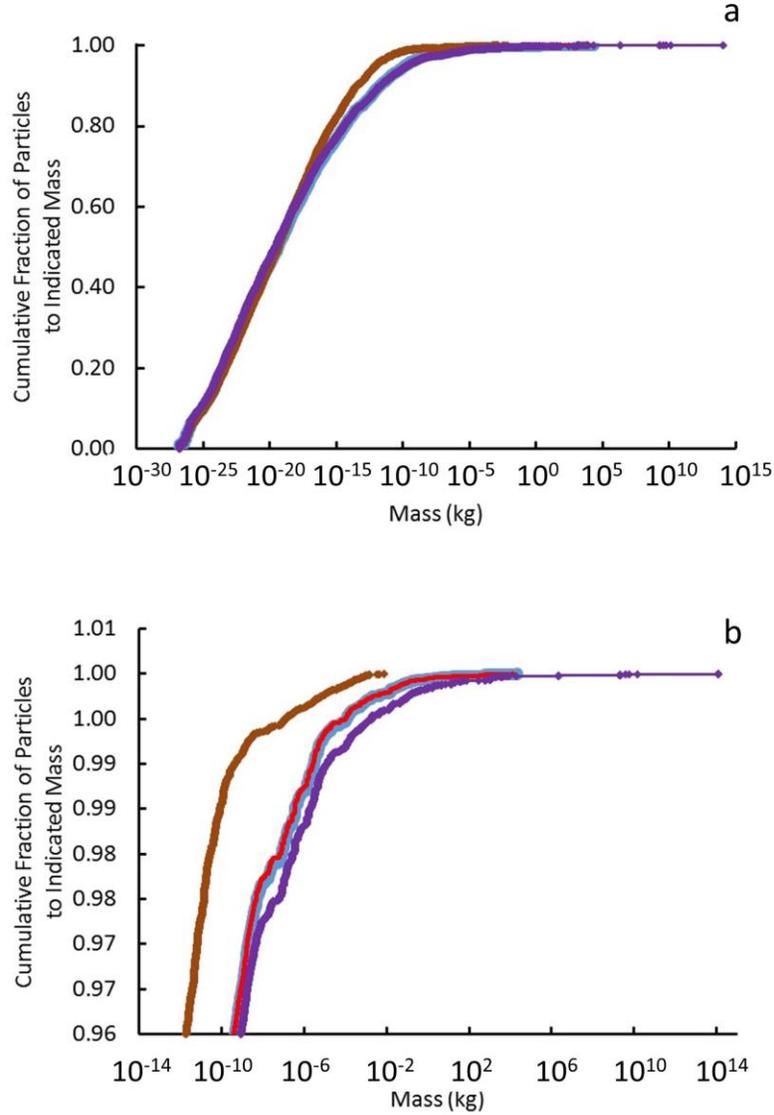

**Figure 2.** Plots of cumulative fraction of particles $F_M$ with mass less than or equal to the indicated mass for (a) all particles and (b) particles in most massive 4% of distribution are shown for baseline values for $\rho_{DM}(t)$ and $\rho_{QN} = 10^{18}$ kg m$^{-3}$. The red and light blue (visible under the red) curves are the results of our aggregation calculations to times $t = 1.8$ million years and 2.4 trillion years respectively with the baseline assumption of $B_o = 10^{12}$ T. The brown and purple curves show sensitivity of the mass distributions to Tatsumi's [16] extremes in surface magnetic field uncertainty: $B_o = 10^{11}$ T and $10^{13}$ T respectively.

The maximum quark-nugget mass is a strong function of $B_o$ and is ~$10^{-2}$ kg, ~$10^4$ kg, and ~$10^{15}$ kg for $B_o = 10^{11}$ T, $10^{12}$ T, and $10^{13}$ T, respectively.

The distribution has the character of aggregations: the great majority of the particles have very little mass and a few have the great majority of the total mass [41]. MQN mass distribution for the baseline $B_o = 10^{12}$ T in Fig. 2 covers nearly 33 orders of magnitude in mass. The <3.5% of

-9-

the particles with mass >10⁻⁹ kg contain all but ~10⁻⁹ % of the total mass. The maximum mass is ~20,000 kg. The average mass is 0.50 kg, and 0.9998 of the total mass is in particles with mass greater than the average. The mass distribution is too extreme to use average mass for comparison with data, as is often done.

The mass distribution for the $B_o = 10^{12}$ T is also presented in Fig. 3 as the fraction of particles in each decadal mass increment, i.e. the fraction with $m_{decade} = \text{Integer}(\log_{10}(m_{QN}))$.

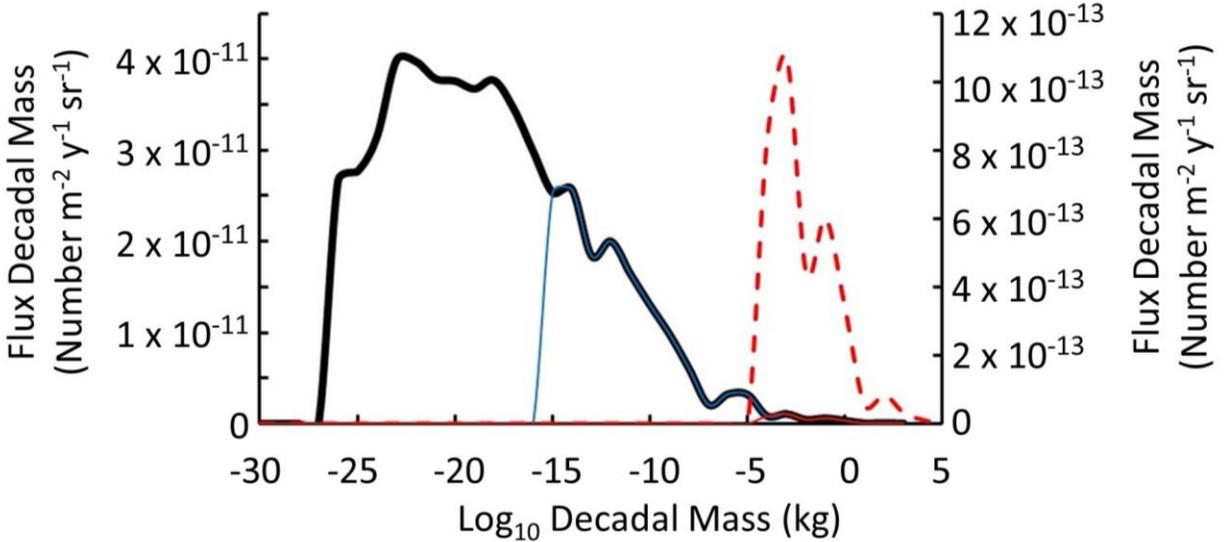

**Figure 3.** Histogram of the flux of quark-nugget masses in each decade of mass from $10^{-27}$ kg to $10^5$ kg for the baseline case of $B_o = 10^{12}$ T and $\rho_{QN} = 10^{18}$ kg/m³, local dark-matter density $\rho_{DM} = 7 \times 10^{-22}$ kg/m³ [4], and local quark nugget velocity $2.5 \times 10^5$ m/s. Solid lines refer to the left axis and dashed lines refer to the right axis. The black line (primary, left) axis shows all quark nuggets and represents the distribution detectable above Earth's atmosphere. The blue line (primary, left) axis represents the distribution that would be detectable in space behind 1 g/cm² of aluminum shielding of the Skylab [36] observations, assuming the quark-nugget's magnetopause [17] dominates its interaction with matter. The solid red line (primary, left) axis and dashed red line (secondary, right) axis represent the distribution detectable after passage through Earth's atmosphere, under the same assumption. About 28% of the quark nuggets incident should be detectable inside Skylab and 0.68% should be detectable at Earth's surface.

We explored the sensitivity of the results to the dark-matter mass density $\rho_{DM}(t)$ by introducing a multiplier to $\rho_{DM}(t)$ given by equation (8). The effect of multipliers 0.1, 0.5, 1.0, 2, and 10 on the final mass distributions is shown in Fig. 4.

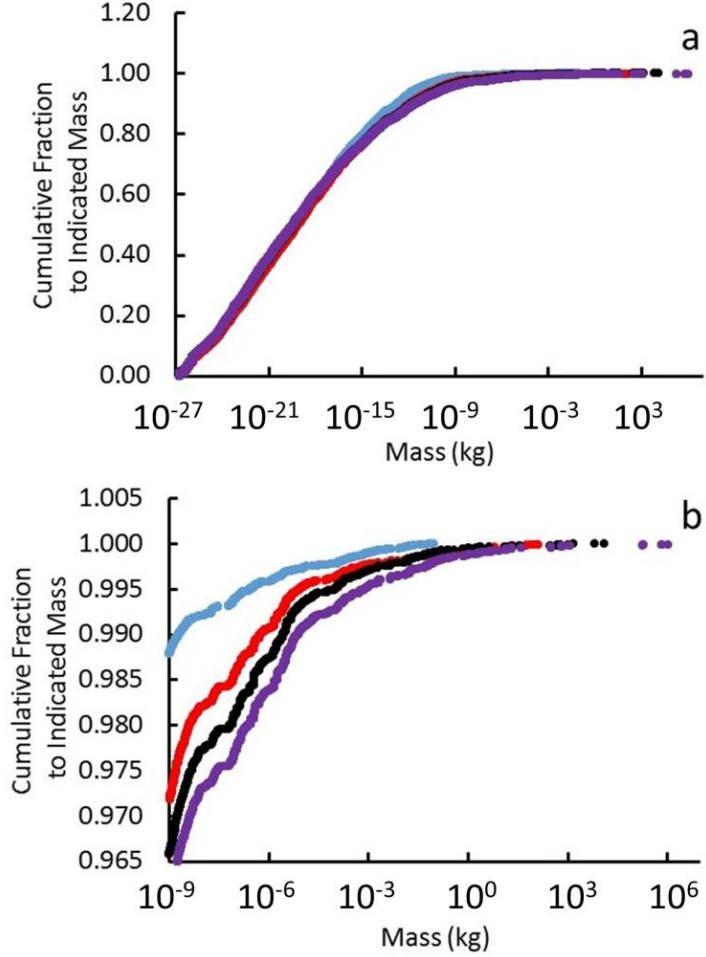

**Figure 4.** Plots of cumulative fraction $F_M$ of particles with mass less than or equal to the indicated mass for four multipliers (0.1 in light blue, 0.5 in red, 1.0 in black, and 2.0 in purple) of dark-matter mass density $\rho_{DM}(t)$ given by equation (8); (a) shows all particles and (b) shows particles with mass $>10^{-9}$ kg.

Variations with changes to $\rho_{DM}$ seem insignificant on the full scale of Fig. 4a. However, comparison of the detailed view in Fig. 4b illustrates a general finding: In the final stages of aggregation, very large masses grow rapidly at the expense of intermediate masses, while very small masses remain relatively unaffected.

The selected view in Fig. 4b also shows that a factor of 2 change in $\rho_{DM}$ can produce a factor of ~100 change in the maximum quark-nugget mass. A factor of 10 reduction in $\rho_{DM}$ can cause a factor of ~$10^5$ reduction in quark-nugget mass. Nevertheless, aggregation robustly increases the quark-nugget masses from their assumed initial mass of ~$10^{-27}$ kg to between 0.1 kg and $10^6$ kg.

These variations with $\rho_{DM}$ are provided to illustrate the effect of the uncertainty in $\rho_{DM}$. Since equations (5) and (6) show aggregation is a function of $\rho_{DM}^{-1} B_o^{-\frac{4}{3}}$, variation in $\rho_{DM}$ gives insight into the effect of the same variation in $B_o^{\frac{4}{3}}$.

As stated in the introduction, uncertainty in $B_o$ drives the uncertainty in the mass distribution. Reducing that uncertainty is the goal of the observations. Then equation (1) can be used to adjust mass distribution when better values of $\rho_{DM}$ and $\rho_{QN}$ are obtained from observations, as illustrated in the Discussion section.

In the next subsection, we will see if this final range of masses is sufficiently large to assure magnetized quark nuggets satisfy the non-interaction requirements for dark matter.

**Effects of MQN mass distributions**

Jacobs, Starkman, and Lynn [12] systematically compare the interaction cross section $\sigma_x$ of unmagnetized quark-nuggets of mass $m_x$ with observations and requirements of dark matter, assuming quark nuggets have a single mass. Their analysis covers large-scale structure constraints (cosmic microwave background and the Bullet Galaxy) from self-interactions and interactions with other baryons and photons, ancient mica observations [32], and results from the Skylab experiment [36]. They assume the cross section for interactions is approximately the geometrical cross section. We adapt their single-mass analysis to the case of MQNs with the computed mass distribution in Fig. 2 and interacting through their magnetic field in vacuum and their magnetopause [17] in a surrounding plasma.

Consider local density of dark-matter $\rho_{DM} = 7 \times 10^{-22}$ kg/m$^3$ [4] and quark nuggets with a single mass $m_{QN}$ and the same velocity $u_{QN}$ relative to a target of identical particles with target number density $n_t$. The interaction cross section $\sigma_x$ is the same for all particles and the event rate per unit volume is $\frac{\Delta n}{\Delta t} = n_{QN} u_{QN} \sigma_x n_t$, in which quark nugget number density $n_{QN} = \frac{\rho_{DM}}{m_{QN}}$. Solving for $\sigma_x/m_{QN}$ gives

$$\frac{\sigma_x}{m_{QN}} = \frac{1}{\rho_{QN} u_{QN} n_t} \frac{\Delta n}{\Delta t}. \tag{10}$$

Similarly, consider a sample of $N_{QN}$ quark nuggets that fill a volume $V$. The number density of quark nuggets is $n_{QN} = \frac{N_{QN}}{V}$. Let the mass, velocity (relative to a target particle), and interaction cross section of the $i^{\text{th}}$ quark nugget be $m_i$, $u_i$, and $\sigma_i$, respectively. Then $V = \frac{\sum_{i=1}^{N_{qn}} m_i}{\rho_{DM}}$ and the event rate per unit volume is

$$\frac{\Delta n}{\Delta t} = \frac{n_t \sum_{i=1}^{N_{qn}}(u_i \sigma_i)}{V} = \frac{n_t \rho_{DM} \sum_{i=1}^{N_{qn}}(u_i \sigma_i)}{\sum_{i=1}^{N_{qn}} m_i}. \qquad (11)$$

The number density $n_{QN}$ of quark nuggets and mean mass $m_{mean}$ are respectively

$$n_{QN} = \frac{\rho_{DM}}{m_{mean}}$$
$$m_{mean} = \frac{\sum_{i=1}^{N_{QN}} m_i}{N_{QN}}. \qquad (12)$$

The approximate expression for a broad mass distribution corresponding to equation (10) is

$$\frac{\sigma_{eff}}{m_{mean}} = \frac{1}{\rho_{DM} u_{eff} n_t} \frac{\Delta n}{\Delta t} = \frac{\sum_{i=1}^{N_{QN}}(u_i \sigma_i)}{u_{eff} \sum_{i=1}^{N_{QN}} m_i}, \qquad (13)$$

in which effective velocity $u_{eff} \approx 3 \times 10^5$ m/s according to Jacobs, *et al.*, and including the ~2.5 × $10^5$ m/s velocity [9] of the solar system about the galactic center for Earth-based observations. Since we have $m_i$, $u_i$ for $N_{QN} = 10^5$, we can compare ($\sigma_{eff}$, $m_{mean}$) with the various criteria Jacobs, *et al.* developed for single-mass quark-nugget dark matter.

As explained in the Methods section, $m_{mean}$ doubles during each generation of 50,000 aggregations of our simulation, which starts each generation with 100,000 particles. Since the self-interaction cross section of equation (4) is proportional to $(m_i + m_j)^{2/3}$, $\frac{\sigma_{eff}}{m_{mean}}$ decreases with each generation and, therefore, with time. Since Fig. 1 shows that the quark-nugget mass distribution at time $t = 1$ ms is very close to its final value, if $\frac{\sigma_{eff}}{m_{mean}}$ criteria are satisfied at $t = 1$ ms, they are sufficiently satisfied at the later times evaluated by Jacobs, *et al.*

Since MQNs are just magnetized Macros, the many interactions of Macros examined by Jacobs, *et al.* that are not affected by the self-magnetic field of the quark nugget are also appropriate to MQNs. The geometric cross section for these non-magnetic interactions varies as $r_{QN}^2$ and the mass varies as $r_{QN}^{-3}$, so $\frac{\sigma_{eff}}{m_{mean}} \propto \frac{1}{r_{QN}} \propto \frac{1}{m_{QN}^{1/3}}$ and the largest value of $\frac{\sigma_{eff}}{m_{mean}}$ is the case with the smallest masses. As shown in Fig. 2, that worst case is $B_o = 10^{11}$ T. Evaluating equation (13) for that worst case gives

$$\max \frac{\sigma_{eff}}{m_{mean}} = \frac{\sum_{i=1}^{N_{QN}}(u_i \sigma_i)}{u_{eff}\sum_{i=1}^{N_{QN}} m_i} = 7 \times 10^{-19} \text{ m}^2 \text{ kg}^{-1} = 7 \times 10^{-18} \text{ cm}^2 \text{ g}^{-1}. \tag{14}$$

The non-magnetic phenomenon with the most stringent scattering requirement evaluated by Jacobs, *et al.* is elastic dark-matter/photon scattering. It requires $\frac{S_{eff}}{m_{mean}} < 4.5 \times 10^{-7}$ cm$^2$ g$^{-1}$ and, as shown by equation (14), is easily satisfied for the worst case MQN distributions.

Equation (13) assumes the cross section $\sigma_i$ depends only on the properties of the i$^{th}$ MQN. Self-interactions are more complicated. MQNs have different masses, so they have different magnetic dipoles. They also have different velocities. The cross section in equation (13) is generalized to $\sigma_{ij}$ in equation (14) and computed from the properties of both the i$^{th}$ and j$^{th}$ MQN as a particle pair. For each aggregation, i$^{th}$ and j$^{th}$ MQNs were randomly chosen from the ensemble of MQNs to evaluate self-interaction with equation (14). The time interval between $t = 1$ ms and 2 ms was chosen as a relevant case because the mass distributions in that time interval are quite similar to those at t = 10 Gyr, as shown in Fig. 1, so they are relevant to the effect of MQN dark matter on the evolution of the universe.

The calculation is also a worst case because the mean MQN mass continues to increase with time and the ratio $\frac{\sigma_{eff}}{m_{mean}}$ decreases with increasing mass. The cross section for magnetic aggregation is used as a surrogate for elastic scattering. The rapidly diminishing magnetic field with increasing distance from a quark nugget assures that the annular cross section for scattering is less than or comparable to the cross section for aggregation. The calculation gives $\frac{\sigma_{eff}}{m_{mean}} = 4.6 \times 10^{-7}$ cm$^2$ g$^{-1}$ and is less than Jacobs, *et al.*'s most conservative 0.04 cm$^2$ g$^{-1}$. Consequently, the calculated mass distributions of MQNs easily satisfy the self-interaction criterion for dark matter.

Protons can also scatter off the magnetic fields of MQNs. The cross section depends on the mass of the MQN and on the velocity of the proton in the rest frame of the MQN. For a representative velocity of $2.5 \times 10^5$ m/s, simulations of proton-MQN scattering in the equatorial plane of the MQN with baseline parameters of $B_o$ and $\rho_{QN}$, gave $\frac{\sigma_x}{m_{QN}} = 2.3 \times 10^{-4}$ cm$^2$ g$^{-1}$ ± 6% and was inversely proportional to proton velocity. Using these results and equation (13) to include the effect of the mass distribution gives $\frac{\sigma_{eff}}{m_{mean}} = 1.9 \times 10^{-4}$ cm$^2$ g$^{-1}$ ± 10% for MQN-baryon scattering at t ≈ 1 ms after the big bang and decreasing with increasing time. Jacobs, *et al.* require $\frac{\sigma_{eff}}{m_{mean}} <$ 0.06 cm$^2$ g$^{-1}$ to be consistent with the observation that dark matter concentrated near galactic centers does not increase gas temperature by collisional heating. Therefore, MQNs comfortably satisfy the requirement in spite of their large magnetic field.

**Effect of escape velocity from galaxy**

In the simulations, each quark nugget's velocity evolved to its final velocity by aggregation with conservation of linear momentum. As discussed with respect to Fig. 4, very large masses grow rapidly at the expense of intermediate masses during aggregation, while very small masses remain relatively unaffected. Therefore, it is not surprising that approximately a third of the quark-nuggets keep their initial high velocity, which exceeds galactic escape velocity. Approximately 35% of the final quark nuggets had total velocity >600 km/s and would not be gravitationally bound within our galaxy. These quark nuggets do not contribute to local dark matter and are dropped from analysis of quark-nugget detections on Earth. However, the 35% of quark nuggets that escape the galaxy have less than 1% of the total MQN mass.

**Direct detection of quark nugget events**

Jacobs, *et al*. found that the null results from the Skylab experiment [36] excluded quark nuggets with a single mass of less than $2 \times 10^{-10}$ kg based on inability to penetrate 0.25 cm of Lexan polycarbonate to make tracks in plastic with >400 MeV cm$^2$ g$^{-1}$ stopping power. The calculation does not include Skylab's 1 g/cm$^2$ aluminum wall. Applying the magnetopause interaction model [17] to quark-nugget passage through the Skylab wall and into the plastic, we find quark-nuggets with mass $\leq 10^{-16}$ kg are effectively shielded from the detector, as shown in Fig. 3. The computed quark-nugget flux $F_j$ in the $j^{th}$ decadal mass increment, which includes all quark nuggets with mass $m_{QN}$ such that $j = \text{Integer}(\log_{10}(m_{QN}))$, is

$$F_j = \rho_{DM} u_{DM} \frac{M_j}{\sum_{j=-27}^{j=5} M_j} \frac{1}{3 \times 10^j} \frac{3.15 \times 10^7}{4\pi}, \quad (15)$$

in which $M_j$ is the total mass in decadal increment $j$, $3 \times 10^j$ is approximately the average quark-nugget mass in $j^{th}$ decade, $3.15 \times 10^7$ is the number of seconds per year, and $4\pi$ sr gives the decadal flux in number m$^{-2}$ y$^{-1}$ sr$^{-1}$.

Jacobs, *et al*. find that the Skylab detector should have been sensitive to unmagnetized quark-nuggets with $\sigma/m_{QN} < 3$ cm$^2$ g$^{-1}$ and rules out single-mass quark nuggets with $m_{QN}$ less than approximately $2 \times 10^{-10}$ kg. The mass distribution for MQNs gives a very different result. Summing flux $F_j$ by decadal mass in Fig. 3 for $j > -16$, i.e. $m_{QN} > 10^{-16}$ kg, gives $1.5 \times 10^{-10}$ m$^{-2}$ y$^{-1}$ sr$^{-1}$ total expected flux of quark nuggets into the Skylab experiment. The experiment had an exposure of about 2 m$^2$ y sr. The expected number of events calculated from the total flux, exposure area and time, and $2\pi$ sr equals ~$2 \times 10^{-9}$ events, which is consistent with the 0 observed. Even if the plastic detector had been outside Skylab, the number of predicted events would have been only $6 \times 10^{-9}$. Skylab-like experiments cannot test the MQN dark-matter hypothesis. Very large areas and long exposure times are essential.

More generally, the mass distribution and number flux are strongly dependent on the surface magnetic field $B_o$, which Tatsumi brackets as $10^{12\pm1}$ T. The corresponding number fluxes of high-velocity (>$10^4$ m/s) quark nuggets are shown in Fig. 5 as a function of $B_o$ for three environments: 1) in space near Earth, 2) on Earth's surface (after slowing down through the magnetopause effect in the atmosphere), and 3) on Earth's surface and depositing sufficient energy/length (i.e. ≥100 MJ/m) to make >3.5 m diameter craters [37] in a peat bog.

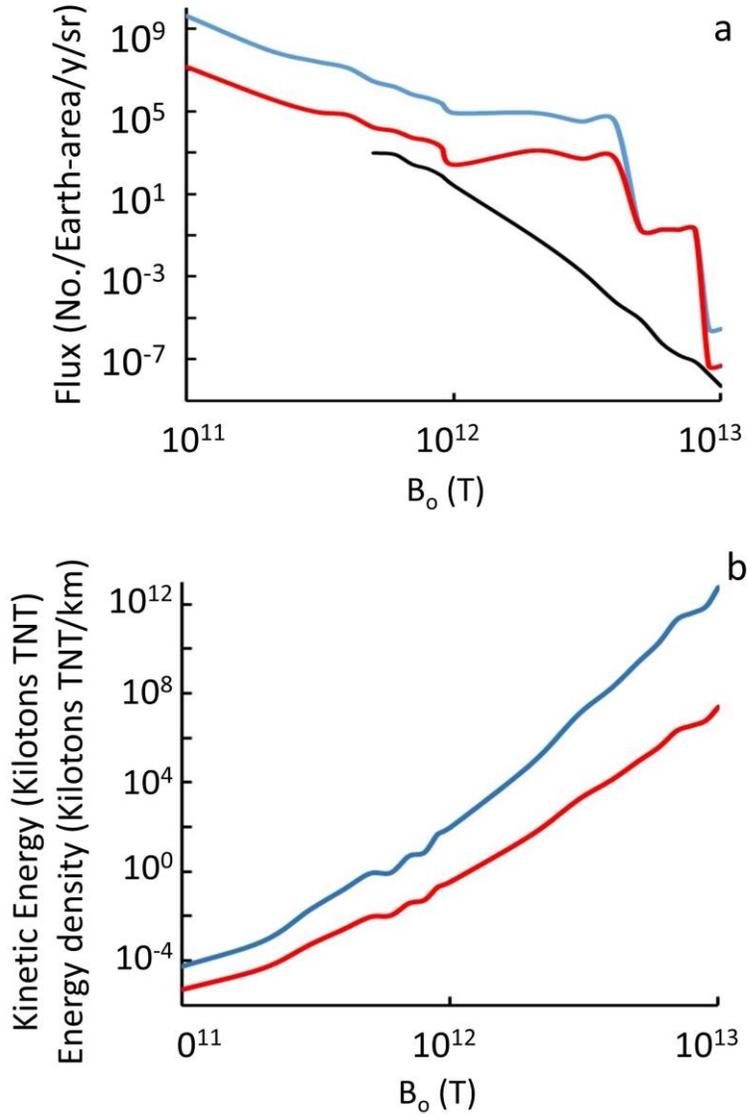

**Figure 5.** a) Quark-nugget number per Earth area ($5 \times 10^{14}$ m$^2$) y$^{-1}$ sr$^{-1}$ of any mass impacting a target above the atmosphere (blue), of mass ≥ $10^{-4}$ kg impacting targets below the atmosphere (red), and of sufficient mass to deposit 100 MJ/m in water (black), as a function of the surface magnetic field $B_o$. Thresholds correspond to space-based targets, acoustically monitored impacts in water, and craters visible from space. b) For the largest mass out of $10^5$ MQNs at the indicated $B_o$, the Kinetic Energy (blue) and Energy Density (red) in the first km of passage through material with mass density 5,500 kg m$^{-3}$ are also shown.

The extremely small number flux of quark-nugget dark matter per Earth-area per year in Fig. 5 means detectors must have very large-area geophysical or planetary targets. The Energy Density (in units of kilotons of TNT per km) is too large for $B_o > 3 \times 10^{12}$ T to have occurred at the indicated flux without having been reported.

The sudden drop in number flux at $B_o \approx 4 \times 10^{12}$ T in Fig. 5 occurs because the aggregation process runs away at that value of $B_o$. Aggregation runaway becomes even more extreme at $B_o \approx 9 \times 10^{12}$ T and creates quark-nuggets with mass $\approx 10^{15}$ kg, as shown in Fig. 6.

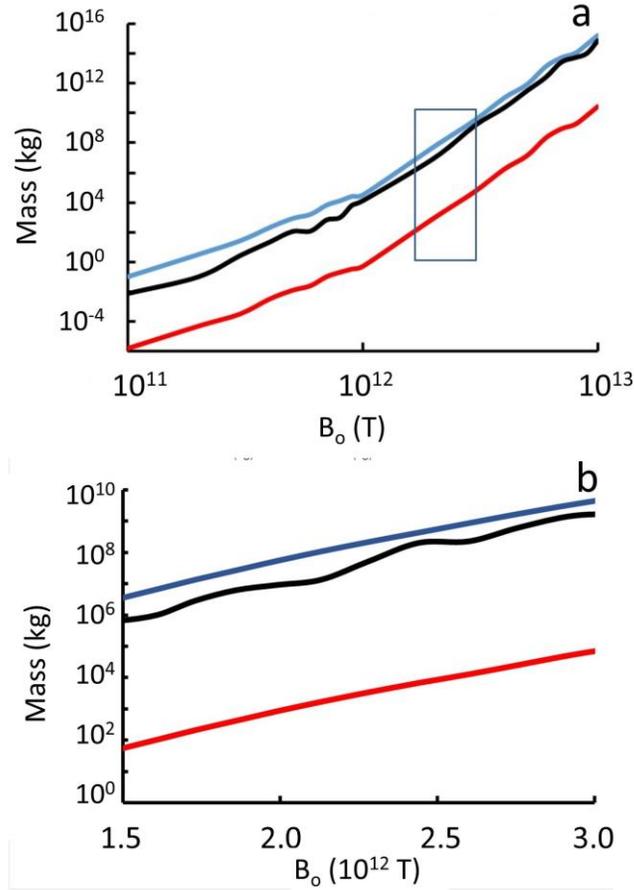

**Figure 6**. Total mass (light blue), most massive quark-nugget (black), and mean mass (red) of simulated quark nuggets with velocity less than escape velocity for the Milky Way; (a) Full range of $B_o$ with box showing most likely range, and (b) Detail view of most likely range of $B_o$ from comparison with observations data in the discussion section.

Computed mass distributions, the mean directed velocity of 250 km/s, and the interstellar mass density of dark matter give representative values of quark-nugget flux by decadal mass and surface magnetic field parameter $B_o$ in Table 1. Results for the entire range of $B_o$ are provided in Supplementary Data: Mass distributions and fluxes by $B_o$ parameter.xlxs. These detailed results can be used to estimate the event rate for future efforts to detect MQNs.

**Table 1:** Representative flux (number in decadal mass interval m$^{-2}$ y$^{-1}$ sr$^{-1}$) by decadal mass for representative values of $B_o$ assuming mean velocity equals 250 km/s and the local mass density of dark matter equals the interstellar value of ~7 × 10$^{-22}$ kg/m$^3$.

| Decadal Mass (kg) | $B_o = 0.1$ ($10^{12}$ T) | $B_o = 0.2$ ($10^{12}$ T) | $B_o = 1.5$ ($10^{12}$ T) | $B_o = 2.0$ ($10^{12}$ T) | $B_o = 3.0$ ($10^{12}$ T) | $B_o = 4.0$ ($10^{12}$ T) |
|---|---|---|---|---|---|---|
| ~ 3 × 10$^{-23}$ | 0 | 0 | 3.5 × 10$^{-16}$ | 2.4 × 10$^{-17}$ | 1.3 × 10$^{-19}$ | 0 |
| ~ 3 × 10$^{-22}$ | 8.4 × 10$^{-8}$ | 1.9 × 10$^{-9}$ | 3.9 × 10$^{-14}$ | 2.7 × 10$^{-15}$ | 1.5 × 10$^{-17}$ | 6.5 × 10$^{-19}$ |
| ~ 3 × 10$^{-21}$ | 1.7 × 10$^{-6}$ | 3.6 × 10$^{-8}$ | 5.5 × 10$^{-13}$ | 3.8 × 10$^{-14}$ | 2.1 × 10$^{-16}$ | 9.6 × 10$^{-18}$ |
| ~ 3 × 10$^{-20}$ | 3.6 × 10$^{-6}$ | 8.2 × 10$^{-8}$ | 1.2 × 10$^{-12}$ | 8.4 × 10$^{-14}$ | 4.6 × 10$^{-16}$ | 2.1 × 10$^{-17}$ |
| ~ 3 × 10$^{-19}$ | 4.0 × 10$^{-6}$ | 8.4 × 10$^{-8}$ | 1.2 × 10$^{-12}$ | 8.3 × 10$^{-14}$ | 4.5 × 10$^{-16}$ | 2.1 × 10$^{-17}$ |
| ~ 3 × 10$^{-18}$ | 4.0 × 10$^{-6}$ | 8.7 × 10$^{-8}$ | 1.2 × 10$^{-12}$ | 8.5 × 10$^{-14}$ | 4.6 × 10$^{-16}$ | 2.1 × 10$^{-17}$ |
| ~ 3 × 10$^{-17}$ | 3.8 × 10$^{-6}$ | 8.5 × 10$^{-8}$ | 1.1 × 10$^{-12}$ | 7.7 × 10$^{-14}$ | 4.2 × 10$^{-16}$ | 1.9 × 10$^{-17}$ |
| ~ 3 × 10$^{-16}$ | 3.4 × 10$^{-6}$ | 7.5 × 10$^{-8}$ | 9.7 × 10$^{-13}$ | 6.7 × 10$^{-14}$ | 3.6 × 10$^{-16}$ | 1.7 × 10$^{-17}$ |
| ~ 3 × 10$^{-15}$ | 2.8 × 10$^{-6}$ | 6.3 × 10$^{-8}$ | 8.2 × 10$^{-13}$ | 5.6 × 10$^{-14}$ | 3.1 × 10$^{-16}$ | 1.4 × 10$^{-17}$ |
| ~ 3 × 10$^{-14}$ | 2.5 × 10$^{-6}$ | 6.2 × 10$^{-8}$ | 8.4 × 10$^{-13}$ | 5.7 × 10$^{-14}$ | 3.1 × 10$^{-16}$ | 1.4 × 10$^{-17}$ |
| ~ 3 × 10$^{-13}$ | 1.9 × 10$^{-6}$ | 4.3 × 10$^{-8}$ | 6.0 × 10$^{-13}$ | 4.2 × 10$^{-14}$ | 2.3 × 10$^{-16}$ | 1.0 × 10$^{-17}$ |
| ~ 3 × 10$^{-12}$ | 1.9 × 10$^{-6}$ | 4.8 × 10$^{-8}$ | 6.5 × 10$^{-13}$ | 4.5 × 10$^{-14}$ | 2.4 × 10$^{-16}$ | 1.1 × 10$^{-17}$ |
| ~ 3 × 10$^{-11}$ | 1.0 × 10$^{-6}$ | 3.4 × 10$^{-8}$ | 5.4 × 10$^{-13}$ | 3.7 × 10$^{-14}$ | 2.0 × 10$^{-16}$ | 9.2 × 10$^{-18}$ |
| ~ 3 × 10$^{-10}$ | 6.3 × 10$^{-7}$ | 2.3 × 10$^{-8}$ | 4.2 × 10$^{-13}$ | 2.9 × 10$^{-14}$ | 1.6 × 10$^{-16}$ | 7.1 × 10$^{-18}$ |
| ~ 3 × 10$^{-9}$ | 2.6 × 10$^{-7}$ | 1.4 × 10$^{-8}$ | 3.2 × 10$^{-13}$ | 2.2 × 10$^{-14}$ | 1.2 × 10$^{-16}$ | 5.5 × 10$^{-18}$ |
| ~ 3 × 10$^{-8}$ | 1.2 × 10$^{-7}$ | 5.6 × 10$^{-9}$ | 2.1 × 10$^{-13}$ | 1.4 × 10$^{-14}$ | 7.9 × 10$^{-17}$ | 3.6 × 10$^{-18}$ |
| ~ 3 × 10$^{-7}$ | 5.5 × 10$^{-8}$ | 2.1 × 10$^{-9}$ | 7.4 × 10$^{-14}$ | 5.1 × 10$^{-15}$ | 2.7 × 10$^{-17}$ | 1.2 × 10$^{-18}$ |
| ~ 3 × 10$^{-6}$ | 7.5 × 10$^{-8}$ | 2.5 × 10$^{-9}$ | 1.2 × 10$^{-13}$ | 8.0 × 10$^{-15}$ | 4.4 × 10$^{-17}$ | 2.0 × 10$^{-18}$ |
| ~ 3 × 10$^{-5}$ | 7.2 × 10$^{-8}$ | 1.8 × 10$^{-9}$ | 1.2 × 10$^{-13}$ | 8.2 × 10$^{-15}$ | 4.5 × 10$^{-17}$ | 2.0 × 10$^{-18}$ |
| ~ 3 × 10$^{-4}$ | 5.0 × 10$^{-8}$ | 7.2 × 10$^{-10}$ | 3.5 × 10$^{-14}$ | 2.6 × 10$^{-15}$ | 1.5 × 10$^{-17}$ | 6.8 × 10$^{-19}$ |
| ~ 3 × 10$^{-3}$ | 5.3 × 10$^{-8}$ | 1.3 × 10$^{-9}$ | 4.5 × 10$^{-14}$ | 3.2 × 10$^{-15}$ | 1.8 × 10$^{-17}$ | 8.3 × 10$^{-19}$ |
| ~ 3 × 10$^{-2}$ | 8.8 × 10$^{-9}$ | 8.1 × 10$^{-10}$ | 1.9 × 10$^{-14}$ | 1.7 × 10$^{-15}$ | 1.0 × 10$^{-17}$ | 4.6 × 10$^{-19}$ |
| ~ 3 × 10$^{-1}$ | 0 | 7.0 × 10$^{-10}$ | 2.6 × 10$^{-14}$ | 1.8 × 10$^{-15}$ | 9.7 × 10$^{-18}$ | 4.4 × 10$^{-19}$ |
| ~ 3 × 10$^{0}$ | 0 | 6.7 × 10$^{-11}$ | 1.9 × 10$^{-14}$ | 1.6 × 10$^{-15}$ | 9.6 × 10$^{-18}$ | 4.6 × 10$^{-19}$ |
| ~ 3 × 10$^{1}$ | 0 | 0 | 4.2 × 10$^{-15}$ | 4.1 × 10$^{-16}$ | 2.9 × 10$^{-18}$ | 1.4 × 10$^{-19}$ |
| ~ 3 × 10$^{2}$ | 0 | 0 | 2.1 × 10$^{-15}$ | 1.9 × 10$^{-16}$ | 6.6 × 10$^{-19}$ | 3.0 × 10$^{-20}$ |
| ~ 3 × 10$^{3}$ | 0 | 0 | 2.1 × 10$^{-15}$ | 1.5 × 10$^{-16}$ | 2.4 × 10$^{-18}$ | 1.0 × 10$^{-19}$ |
| ~ 3 × 10$^{4}$ | 0 | 0 | 3.5 × 10$^{-16}$ | 0 | 0 | 0 |
| ~ 3 × 10$^{5}$ | 0 | 0 | 1.4 × 10$^{-15}$ | 0 | 0 | 0 |
| ~ 3 × 10$^{6}$ | 0 | 0 | 0 | 1.5 × 10$^{-16}$ | 2.7 × 10$^{-19}$ | 1.8 × 10$^{-20}$ |
| ~ 3 × 10$^{7}$ | 0 | 0 | 0 | 0 | 0 | 0 |
| ~ 3 × 10$^{8}$ | 0 | 0 | 0 | 0 | 1.3 × 10$^{-19}$ | 0 |
| ~ 3 × 10$^{9}$ | 0 | 0 | 0 | 0 | 1.3 × 10$^{-19}$ | 2.4 × 10$^{-20}$ |
| ~ 3 × 10$^{10}$ | 0 | 0 | 0 | 0 | 0 | 1.2 × 10$^{-20}$ |
| **All** | **3.21 × 10$^{-5}$** | **7.53 × 10$^{-7}$** | **1.1 × 10$^{-11}$** | **7.7 × 10$^{-13}$** | **4.2 × 10$^{-15}$** | **1.92 × 10$^{-16}$** |

For $B_o > 1.5 \times 10^{12}$ T, Table 1 shows gaps in the distribution of flux as a function of mass. The continuous distribution breaks into two or even three isolated distributions. The effect is aggregation runaway. Since the cross section for aggregation in equation 4 is larger for two large-mass MQNs than it is for one large-mass and one small-mass MQN, the large mass aggregations preferentially aggregate and leave gaps in the distribution. Aggregation also reduces MQN velocity, so the velocity difference in the denominator of equation 4 also contributes to the runaway effect. Runaway explains why the flux rates decrease dramatically for large $B_o$.

**Discussion**

Tatsumi [16] found that quark nuggets could exist as a ferromagnetic fluid held together by the strong nuclear force and have a surface magnetic field between $10^{11}$ T and $10^{13}$ T. We have explored the consequences of his theory as applied to dark matter. In this paper, the direct simulation Monte Carlo method shows magnetized quark nuggets (MQNs) that formed with baryon number $A = 1$ when the universe is at ~100 MeV, approximately $t = 65$ μs after the big bang, aggregate under the influence of their self-magnetic fields. We find that aggregation dominates decay by the weak interaction and produces mass distributions, as a function of the $B_o$ parameter, that satisfy requirements [12] for dark matter as early as 1 ms after the universe forms.

Previous searches for dark matter have been based on single-mass analysis. Mass distributions computed from first principles help design precision tests of the MQN dark matter hypothesis. The mass distributions reported in this paper and the magnetopause interactions described in Ref. 17 make the MQN hypothesis for dark matter testable. All quark nuggets have mass density somewhat greater than nuclear density ~ $10^{18}$ kg/m$^3$, so their geometric cross section $\sigma_o$ is very small. The energy deposited per unit length in passage through matter is correspondingly very small. However, each MQN has a very large magnetic field that is compressed by the particle pressure of ionized matter (plasma) streaming into its rest frame with relative velocity $v$. The particle pressure is balanced by the compressed magnetic field pressure at its magnetopause, which has a cross section $\sigma_m$ that is much larger than its geometric cross section $\sigma_o$ [17].

$$\sigma_m = \left( \frac{2B_o^2}{\mu_0 K \rho_p v^2} \right)^{\frac{1}{3}} \sigma_o, \qquad (16)$$

in which $\mu_o$ = permeability of free space, $K \sim 1$, and $\rho_p$ = the local plasma density. The relative velocity $v$ is on the order of the velocity of the Earth's motion about the galactic center and through the dark-matter halo, i.e. ~250 km/s. At that velocity, incoming matter is quickly heated to the plasma state upon impact, similar to how a meteor creates a plasma passing through atmosphere. The decelerating force $F_e$ from the drag on the magnetopause is

$$F_e = m \frac{d^2 x}{dt^2} = -K \sigma_m \rho_p v^2. \qquad (17)$$

Integrating equation (17) twice with the cross section from equation (16) gives the distance $x_{max}$ at which velocity $v = 0$ for an MQN with mass $m$, mass density $\rho_{QN}$, and initial velocity $v_o$ for a given $B_o$:

$$x_{max} = \left( \frac{3\mu_o \rho_{QN}^2 v_o^2 m}{\pi K^2 \rho_p^2 B_o^2} \right)^{\frac{1}{3}}. \qquad (18)$$

By definition, the force $F_e$ in equation (17) equals the energy deposition per unit length. For example, Fig. 6 shows the maximum mass in a collection of 100,000 MQNs in the simulation as a function of $B_o$. Fig. 5b shows the kinetic energy (blue) and energy/km deposited (red) by the maximum mass as a function of $B_o$. The simulation also gives the number flux as a function of mass and $B_o$, as illustrated in Table 1. Combining this information gives the approximate event rate for impacts above a minimum energy deposited per unit length as a function of $B_o$. The fact that Megaton-TNT/km impacts are not observed every century or so makes $B_o > 3 \times 10^{12}$ T very unlikely. We, therefore, exclude $B_o > 3 \times 10^{12}$ T.

In addition, Fig. 5a shows that etched-plastic targets in space without any shielding (blue) would require between ~$4 \times 10^9$ m$^2$ and ~$3 \times 10^4$ m$^2$ area to detect one event per year for $B_o$ between $4 \times 10^{12}$ T and $10^{11}$ T, respectively. Such large detectors in space are impractical.

Before the MQN mass distribution was calculated, Ref. 17 proposed acoustically monitoring MQN impacts in water at the Great Salt Lake, Utah, USA, as a possible sensor for MQNs. Energy absorption in the atmosphere prevents detection of MQN masses $m < 10^{-4}$ kg [17]. However, energy/length deposited in water of density ~ 1 kg/m$^3$ for $B_o \sim 10^{12}$ T is ~ 10 kJ/m for $m = 10^{-4}$ kg and is ~ 100 MJ/m for $m \sim 10$ kg. Such a large energy density would create large amplitude acoustic waves that would be easily detected by an array of hydrophones. Short-range testing, reported in Ref. 17, indicated that the background noise and sound propagation through the water would allow detection over most of the Great Salt Lake. When the full system was deployed, long-range testing included absorption and interference effects in the water and material beneath the lake. These effects are not yet fully understood but apparently limited the range to ~ 3 km. Therefore, the array was sensitive to MQN impacts with mass $\geq 10^{-4}$ kg within a ~$3 \times 10^7$ m$^2$ area, which is about 30 times the detector area of IceCube Neutrino Observatory at the South Pole. Weather limited observations to ~90 days per year. The system was calibrated with line explosives extending from the surface to the bottom and at various distances to obtain the signature of an MQN impact. No events were recorded with the distinctive signature in 90 days of good observations in 2019.

Mean mass is often used for dark-matter candidates. The Poisson probability of obtaining $k$ events with the expected mean rate $\lambda$ is given by

$$P(\lambda, k) = \frac{e^{-\lambda} \lambda^{-k}}{k!}. \qquad (19)$$

The signature from an MQN impact is so far above background for $m \geq 10^{-4}$ kg that the background is effectively 0. For the mean-mass analysis, the expected mean rate is

$$\lambda = \frac{2\pi \rho_{DM} v (Area)}{m_{mean}}, \quad (20)$$

in which $2\pi$ = solid angle of incident MQNs, interstellar dark-matter density $\rho_{DM} = 7 \times 10^{-22}$ kg/m$^3$, mean incident velocity $v = 250$ km/s, detector $Area = 3 \times 10^7$ m$^2$, and $m_{mean}$ is the mean MQN mass. Probability $P$ from equation (19) for null result $k = 0$ equals $10^{-144}$, 0.01, and 0.25 for $m_{mean} = 10^{-4}$ kg, $7.2 \times 10^{-3}$ kg, and 0.025 kg respectively. Therefore, the mean-mass analysis implies, $m_{mean} \leq 7.2 \times 10^{-3}$ kg is excluded at the 99% confidence level. Fig. 6a gives $m_{mean}$ versus $B_o$. Excluding $m_{mean} \leq 7.2 \times 10^{-3}$ kg excludes $B_o \leq 2.5 \times 10^{11}$ T.

The same analysis for distributed mass has been done from the results in Table 1 and can also be done with less precision from in Fig. 5a (red). For $B_o = 1 \times 10^{11}$ T, $2 \times 10^{11}$ T, and $3 \times 10^{11}$ T, $\lambda = $ 5.2, 0.17, and 0.036, respectively, and $P$ for $k = 0$ equals 0.005, 0.84, and 0.96, respectively. Therefore, analysis of the data with the computed mass distributions excludes distributions with $B_o \leq 1 \times 10^{11}$ T with a confidence level of 99.5% from Poisson statistics. Comparison with the mean-mass analysis illustrates the importance of the mass distribution.

The strong dependence of $P$ with $B_o$ implies that a much larger area detector and/or much longer time is required to investigate the remaining parameter space of $1 \times 10^{11}$ T $< B_o \leq 3 \times 10^{12}$ T. Asteroids provide some possibility of detection over a much longer time because they should have been accumulating MQNs since the solar system was formed ~4 Gyr ago. The MQN mass accumulated in an asteroid is conservatively estimated by assuming an asteroid of radius $r_a$ stops incident MQNs with range $x_{max}$ if $r_a \geq x_{max}$. The maximum MQN mass $m_{max}$ that will be absorbed by an asteroid with radius $r_a$ is given by solving equation (18) for $m = m_{max}$ with $x_{max} = r_a$.

Table 1 provides the number flux (number m$^{-2}$ y$^{-1}$ sr$^{-1}$) for MQN mass $m$ for a given $B_o$. Multiplying each number flux by the associated decadal mass gives the mass flux by decadal mass. If all dark matter is composed of MQNs, the accumulated MQN mass in an asteroid is estimated as follows:
1. Sum decadal mass fluxes for masses less than $m_{max}$,
2. multiply the sum by cross sectional area $\pi r_a^2$,
3. multiply that by 5.56 sr to include MQNs incident from all directions [38], and
4. multiply that by 4 Gyr for the accumulation time.

For $B_o = 1.5 \times 10^{12}$ T, the total MQN mass $m_{QN\_total} \approx 2 \times 10^{-11} r_a^{4.33}$ kg for $r_a \geq 100$ m. For example, $m_{QN\_total} \sim 0.01$ kg for $r_a = 100$ m and $m_{QN\_total} \sim 200$ kg for $r_a = 1000$ m.

MQNs accumulated inside an asteroid would be much more detectable if they aggregate into a large MQN. However, aggregation inside an asteroid is much more complicated than it is in free space. The attractive force (in the point-dipole approximation) between two identical magnetic dipoles [47] with magnetic moment $m_m$ separated by a distance x is

$$F_m = \frac{3\mu_o m_m^2}{2\pi x^4}. \quad (21)$$

The $1/x^4$ dependence assures the magnetic force dominates gravity only over short distances, so both forces have to be modeled with realistic distributions of MQN impacts as a function of asteroid size and with realistic resistance of MQN diffusion through asteroid material. Such a simulation is beyond the scope of this paper, but it should indicate if asteroid mining may provide a method to collect MQN dark matter or if such an attempt would be another null result.

Null results may be insufficient to motivate a very difficult systematic search for MQNs. In the past, episodic observations have tested new theories before investments were made for systematic study, e.g. General Relativity was first tested with the anomalous perihelion of Mercury and then with the bending of light near the sun in an eclipse. The results motivated more systematic studies with gravitational red shifts and, a century later, with gravitational waves. Consequently, we looked for episodic opportunities.

Terrestrial craters caused by non-meteorite impacts offer larger areas and longer observation times. In recent years, NASA-investigated, non-meteorite impacts have been reported in the press approximately once per year. Ref. 17 catalogues three such events:
"A12-m diameter crater occurred at 11:05 PM, September 6, 2014, near Managua, Nicaragua [Cooke, W. Did a meteorite cause a crater in Nicaragua? http://blogs.nasa.gov/Watch_the_Skies/2014/09/08/did-a-meteorite-cause-a-crater-in-nicaragua/ and http://www.cnn.com/2014/09/08/tech/innovation/nicaragua-meteorite/, (2014) (Date of access: 24/06/2017)]. An event occurred on July 4, 2015, at the Salty Brine Beach in Rhode Island, USA [Shapiro, E., Cathcart, C. & Donato, C. Bomb squad, ATF investigating mysterious explosion at Rhode Island beach. http://abcnews.go.com/US/explosion-report-prompts-evacuation-rhode-island-beach/story?id=32384143, (2015) (Date of access: 24/06/2017)]. Finally, an event occurred on February 6, 2016, in Tamil Nadu, India [Hauser, C. That wasn't a meteorite that killed a man in India, NASA says. http://www.nytimes.com/2016/02/10/world/asia/that-wasnt-a-meteorite-that-killed-a-man-in-india-nasa-says.html?_r=0, (2016) (Date of access: 24/06/2017)]." [17]. (Each link was also accessed on 15/04/2020.)

These reports occur approximately once per year, as illustrated by these three examples within 3 years. Monitoring peat bogs for non-meteorite impact craters was also proposed in Ref. 17 to link non-meteorite craters to MQN impacts. Irish peat bogs offer up to $3 \times 10^8$ m$^2$ area witness plates that preserve impact craters for 100 to 1000 years, depending on the size of the impacting MQN and the value of $B_o$. One 3.5-m diameter crater from an impact in 1985 has been excavated to the bedrock. The results are consistent with a nearly vertical impact of a 10 +/- 7 kg MQN. The error bar arises from the uncertainty in $B_o$ for $4 \times 10^{11}$ T $\leq B_o \leq 3 \times 10^{12}$ T in Fig. 5a (black). The investigation confirmed the crater is consistent with an MQN with five points of comparison between theory and observation. The non-meteorite crater provided the first--of many needed--results supporting the MQN hypothesis for dark matter, and narrowed the range of allowed $B_o$ [37].

The results from non-meteorite craters motivate finding a systematic way to test the MQN hypothesis with the largest possible detector area. The largest accessible area for real-time MQN searches is Earth's magnetosphere. An experiment to systematically explore the MQN dark-matter hypothesis with a three-satellite constellation at 51,000 km altitude is described in Ref. 38. MQNs experience a net torque as they decelerate during passage through Earth's magnetosphere, ionosphere, and troposphere. The angular acceleration gives them MHz frequencies before they emerge into the magnetosphere where they can be detected by characteristic Doppler-shifted, radiofrequency emissions from their rotating magnetic dipole [38]. Thee MQN mass distributions presented in this paper are essential for planning such a systematic study.

**Methods**

The simulation begins when the temperature $T_{MeV}$ ~ 100 MeV, which occurs at ~65 μs after the Big Bang. We start with $10^5$ particles, as recommended by Kruis, Maisels, and Fissan [41], to provide adequate statistics. The initial speeds of the $10^5$ particles were generated to fit a Maxwell-Boltzmann distribution with temperature $T = 100$ MeV. Their velocity vectors were generated by random selection of velocity unit vectors.

A representative co-moving (i.e. expanding as the universe expands) volume $V$ contains all the masses $m_i$ in the simulation for a generation, with index $i$ varying from 0 to $n$-1 for $n$ particles. Equation (8) provides the dark matter density $\rho_{DM}$ versus time.

The DSMC method for computing aggregations [41] requires the aggregating collision rate $\beta_{ij}^V$ for each particle pair ($i$, $j$), with mass $m_i$ and $m_j$, respectively, and velocity $\vec{u}_i$ and $\vec{u}_j$, respectively:

$$\beta_{ij}^V = \frac{\sigma_{ij} |\vec{u}_i - \vec{u}_j|}{V} \quad . \tag{22}$$

Each $\beta_{ij}^V$ is based on the aggregation cross section $\sigma_{ij}$, which is given by equation (4), relative velocity $|\vec{u}_i - \vec{u}_j|$, and the volume V of the simulation.

$$V = \frac{\sum_{i=0}^{n-1} m_i}{\rho_{DM}} \tag{23}$$

The sum $S_i$ = the sum of all possible collision rates for aggregations with the i$^{th}$ particle and is the collision kernel for the i$^{th}$ particle:

$$S_i = \sum_{l=0, l \neq i}^{n-1} \beta_{il}^V \tag{24}$$

$S$ is the sum of all $S_i$ for the next aggregation event.

$$S = \sum_{l=0}^{n-1} S_l \quad . \tag{25}$$

Since $S$ is the sum of all possible collision rates $\beta_{ij}^V$ ordered by $i$ and then for all $j$ associated with that $i$ before proceeding to the next $i$, every collision pair ($i$, $j$) is represented by a value between 0 and $S$. The numeric interval between ($i$, $j$) entries is proportional to the probability of their

-23-

aggregating. To randomly select a pair $(i, j)$ for collision, a random number $R$ is chosen in the interval 0 to 1. Index $i\_particle$ is chosen for aggregation by solving

$$S_{i\_particle-1} \leq RS \leq S_{i\_particle} . \tag{26}$$

Then Index $j\_particle$ is chosen for aggregation by solving

$$S_{i\_particle-1} + \sum_{l=0, l \neq i\_particle}^{j\_particle-2} \beta^V_{i\_particle,l} \leq RS \leq S_{i\_particle-1} + \sum_{l=0, l \neq i\_particle}^{j\_particle-1} \beta^V_{i\_particle,l} . \tag{27}$$

The process, therefore, produces $(i, j)$ pairs for aggregation in proportion to their probability of aggregating.

To record the aggregation of the i_particle and j_particle, the mass $m_i$ is replaced by the mass $m_i + m_j$, the velocity components of the new $i$th particle are calculated from conservation of linear momentum, the $j$th particle is removed from inventory (reducing the total number of particles by 1), the $\beta_{ij}^V$ terms are subtracted from each $S_i$ to update it and the new value of $S$ is calculated.

The process repeats for 50,000 aggregations, which is one generation. At the end of a generation, the aggregation process has reduced the number of particle by half and the mean mass $m_{mean}$ of the 50,000 remaining particles has doubled.

To keep enough particles, we double the volume and "incorporate" the 50,000 particles from the added volume to restore the particle count to 100,000. Each of the original particles is duplicated with a twin that has the same mass, speed, and kinetic energy. Since elastic collisions are also occurring, the direction of the velocity vector of a particle and the direction of the velocity vector of its twin in the added volume are uncorrelated. Therefore, the directions of the velocities are randomized for the new particles. This randomization is also necessary to prevent division by zero in the calculation of elapsed time, which is inversely proportional to equation (22).

The mean simulated time $\delta_{g,k}$ required for one aggregation is the inverse of half the sum of all collision kernels S in equation (25):

$$\delta_{gk} = \frac{2}{S} . \tag{28}$$

Its expanded form is derived by combining equation (28) with equations (4) and (22) through (25). The result is

$$\delta_g = \sum_{k=0}^{\frac{n}{2}-1} \delta_{g,k} = \sum_{k=0}^{\frac{n}{2}-1} \frac{2}{\rho_{DM,g,k}} \left( \frac{2\mu_o \rho_{QN}^2}{9\pi^{1/2} B_o^2} \right)^{2/3} \left\{ \frac{\sum_{i=0}^{i=n-1-k} m_i}{\sum_{i=0}^{i=n-1-k} \sum_{\substack{j=0 \\ j\neq i}}^{j=n-1-k} \frac{(m_i + m_j)^{\frac{1}{3}}}{|\vec{u}_i - \vec{u}_j|^{\frac{1}{3}}}} \right\}, \qquad (29)$$

which is the expanded form of equation (5).

## Data Availability

All final analyzed data generated during this study are included in this published article plus the accompanying spreadsheet in Supplementary Data: Mass distributions and fluxes by $B_o$ parameter.xlxs.

## Acknowledgements

We gratefully acknowledge S. V. Greene for first suggesting that quark nuggets might explain the geophysical evidence that initiated this research (she generously declined to be a coauthor) and Jesse Rosen for editing the manuscript and suggesting many improvements for clarity.

**Author Contributions**

J. P. V. was lead physicist and principal investigator. He developed computer program to calculate the quark-nugget mass distribution, analyzed the results, wrote the paper, and prepared the figures, and revised the paper to incorporate improvements from the other author and reviewers.

I. M. S. was physicist and lead theorist. He provided the critical function of dark-matter density versus time and temperature, consistent with the standard $\Lambda$CDM cosmology model, and many improvements to the paper.

T. S. was contributing physicist. He provided insight and analysis about the decay of singlets from the weak interaction, which our results show is overcome by aggregation, and about the self-magnetic field of the $A = 1$ singlet, which is ~ $10^{12}$ T and is consistent with the aggregation process.

A. P. V. was lead physicist for the Red Team critique of the paper and provided many improvements to the paper through his reviews and suggestions.

B. A. U. was contributing physicist and provided useful suggestions on how to detect MQNs, including looking for them in asteroids.

**Additional Information**

Competing Financial Interests

The authors declare that there are no competing financial interests.

Figure Legends

**Figure 1.** Snapshots of the masses at various times $\delta t$ after the beginning of the simulation at $t = 65$ μs and for $B_o = 10^{12}$ T. Distribution changes from a single mass = $1.6 \times 10^{-27}$ kg at $\delta t = 0$ to an increasingly broad mass distribution with increasing time.

**Figure 2.** Plots of cumulative fraction of particles $F_M$ with mass less than or equal to the indicated mass for (a) all particles and (b) particles in most massive 4% of distribution are shown for baseline values for $\rho_{DM}(t)$ and $\rho_{QN} = 10^{18}$ kg m$^{-3}$. The red and light blue (visible under the red) curves are the results of our aggregation calculations to times $t = 1.8$ million years and 2.4 trillion years respectively with the baseline assumption of $B_o = 10^{12}$ T. The brown and purple curves show sensitivity of the mass distributions to Tatsumi's [16] extremes in surface magnetic field uncertainty: $B_o = 10^{11}$ T and $10^{13}$ T respectively.

**Figure 3.** Histogram of the flux of quark-nugget masses in each decade of mass from $10^{-27}$ kg to $10^5$ kg for the baseline case of $B_o = 10^{12}$ T and $\rho_{QN} = 10^{18}$ kg/m$^3$, local dark-matter density $\rho_{DM} = 7 \times 10^{-22}$ kg/m$^3$, and local quark nugget velocity $2.5 \times 10^5$ m/s. Solid lines refer to the left axis and dashed lines refer to the right axis. The black line (primary, left) axis shows all quark

nuggets and represents the distribution detectable above Earth's atmosphere. The blue line (primary, left) axis represents the distribution that would be detectable in space behind 1 g/cm$^2$ of aluminum shielding of the Skylab [36] observations, assuming the quark-nugget's magnetopause [17] dominates its interaction with matter. The solid red line (primary, left) axis and dashed red line (secondary, right) axis represent the distribution detectable after passage through Earth's atmosphere, under the same assumption. About 28% of the quark nuggets incident should be detectable inside Skylab and 0.68% should be detectable at Earth's surface.

**Figure 4.** Plots of cumulative fraction $F_M$ of particles with mass less than or equal to the indicated mass for four multipliers (0.1 in light blue, 0.5 in red, 1.0 in black, and 2.0 in purple) of dark-matter mass density $\rho_{DM}(t)$ given by equation (8); (a) shows all particles and (b) shows particles with mass $>10^{-9}$ kg.

**Figure 5.** a) Quark-nugget number per Earth area ($5 \times 10^{14}$ m$^2$) y$^{-1}$ sr$^{-1}$ of any mass impacting a target above the atmosphere (blue), of mass $\geq 10^{-4}$ kg impacting targets below the atmosphere (red), and of sufficient mass to deposit 100 MJ/m in water (black), as a function of the surface magnetic field $B_o$. Thresholds correspond to space-based targets, acoustically monitored impacts in water, and craters visible from space. b) For the largest mass out of $10^5$ MQNs at the indicated $B_o$, the Kinetic Energy (blue) and Energy Density (red) in the first km of passage through material with mass density 5,500 kg m$^{-3}$ are also shown.

**Figure 6**. Total mass (light blue), most massive quark-nugget (black), and mean mass (red) of simulated quark nuggets with velocity less than escape velocity for the Milky Way; (a) Full range of $B_o$ with box showing most likely range, and (b) Detail view of most likely range of $B_o$ from comparison with observations data in the discussion section and other observations being prepared for publication.

Table Legend

**Table 1:** Representative flux (number in decadal mass interval m$^{-2}$ y$^{-1}$ sr$^{-1}$) by decadal mass for representative values of $B_o$ assuming mean velocity equals 250 km/s and the local mass density of dark matter equals the interstellar value of $\sim 7 \times 10^{-22}$ kg/m$^3$.

Supplementary Information for

**Mass distribution of magnetized quark-nugget dark matter and comparison with requirements and observations**

J. Pace VanDevender, Ian Shoemaker, T. Sloan, Aaron P. VanDevender, Benjamin A. Ulmen

**Supplementary Note: Quark-nugget research summary**

Macroscopic quark nuggets [7] are also called strangelets [8], nuclearites [9], AQNs [10], slets [11], and Macros [12]. The theory of quark nuggets by Witten [7] indicates quark-nuggets are in an ultra-dense, color-flavor-locked (CFL) phase [18] of quark matter. Steiner, *et al.* [19] showed that the ground state of the CFL phase is color neutral and that color neutrality forces electric charge neutrality, which minimizes electromagnetic emissions. However, Xia, *et al.* [11] found that quark depletion causes the ratio $Q/A$ of electric charge $Q$ to baryon number $A$ to be non-zero and varying at $Q/A \sim 0.32\, A^{-1/3}$ for $3 < A < 10^5$. In addition to this core charge, they find that there is a large surface charge and a neutralizing cloud of charge to give a net zero electric charge for sufficiently large $A$. So quark nuggets with $A \gg 1$ are both dark and very difficult to detect with astrophysical observations.

Witten and Xia, *et al.* also showed their density should be somewhat larger than the density of nuclei, and their mass very large, even the mass of a star. Large quark nuggets are predicted to be stable [7, 8, 18, 20] with mass between $10^{-8}$ kg and $10^{20}$ kg within a plausible but uncertain range of assumed parameters of quantum chromodynamics (QCD) and the MIT bag model with its inherent limitations [21].

Although Witten assumed a first-order phase transition formed quark nuggets, Aoki, *et al.*[22] showed that the finite-temperature QCD transition that formed quark nuggets in the hot early universe was very likely an analytic crossover, involving a rapid change as the temperature varied, but not a real phase transition. Recent simulations by T. Bhattacharya, *et al*. [23] support the crossover process.

A combination of quark nuggets and anti-quark nuggets have also been proposed within constraints imposed by observations of neutrino flux [24]. Zhitnitsky [10] proposed that Axion Quark Nuggets (AQN) that forms quark and anti-quark nuggets generated by the collapse of the axion domain wall network. Although the model relies on the hypothetical particle that is a proposed extension of the Standard Model to explain CP violation, it appears to explain a wide variety of longstanding problems and leads to quark and anti-quark nuggets with a narrow mass distribution at ~1 kg [25]. Atreya, *et al*. [26] also found that CP-violating quark and anti-quark scatterings from moving Z(3) domain walls should form quark and anti-quark nuggets, regardless of the order of the quark-hadron phase transition.

Experiments by A. Bazavov, *et al*. [27] at the Relativistic Heavy Ion Collider (RHIC) have provided the first indirect evidence of strange baryonic matter. Additional experiments at RHIC may determine whether the process is a first order phase transition or the crossover process. In either case, quark nuggets could have theoretically formed in the early universe.

In 2001, Wandelt, *et al*. [13] showed that quark nuggets meet all the theoretical requirements for dark matter and are not excluded by observations when the stopping power for quark nuggets in the materials covering a detector is properly considered and when the average mass is $>10^5$ GeV (~$2 \times 10^{-22}$ kg). In 2014, Tulin [15] surveyed additional simulations of increasing sophistication and updated the results of Wandelt, *et al*. The combined results help establish the allowed range and velocity dependence of the strength parameter and strengthen the case for quark nuggets. In 2015, Burdin, *et al*. [28] examined all non-accelerator candidates for stable dark matter and also concluded that quark nuggets meet the requirements for dark matter and have not been excluded experimentally. Jacobs, Starkman, and Lynn [12] found that combined Earth-based, astrophysical, and cosmological observations still allow quark nuggets of mass 0.055 to $10^{14}$ kg and $2 \times 10^{17}$ to $4 \times 10^{21}$ kg to contribute substantially to dark matter. The large mass means the number per unit volume of space is small, so detecting them requires a very large-area detector.

These studies did not consider an intrinsic magnetic field within quark nuggets. However, Tatsumi [16] has shown that the lowest-energy configuration of a quark nugget is a ferromagnetic liquid held together by strong nuclear forces. He calculates the value of the magnetic field at the surface of a quark-nugget core inside a magnetar to be $10^{12\pm1}$ T, which is large compared to expected values for the magnetic field at the surface of a magnetar star with a quark-nugget core. For a quark nugget of radius $r_{QN}$ and a magnetar of radius $r_s$, the magnetic field scales as $(r_{QN}/r_s)^3$. Therefore, the surface magnetic field of a magnetar is smaller than $10^{12}$ T because $r_s > r_{QN}$. Since quark-nugget dark matter is bare, the surface magnetic field of what we wish to detect is $10^{12\pm1}$ T.

Although the cross section for interacting with dense matter is greatly enhanced [17] by the magnetic field which falls off as radius $r_{QN}^{-3}$, the collision cross section is still many orders of magnitude too small to violate the collision requirements [12, 13, 15, 28] for dark matter and will be discussed below.

Chakrabarty [29] showed that the stability of quark nuggets increases with increasing magnetic field $\leq 10^{16}$ T, so the large self-field described by Tatsumi should enhance their stability. Ping, *et al.*[30] showed that magnetized quark nuggets should be absolutely stable with the newly-developed equivparticle model, so the large self-field described by Tatsumi should ensure that they will not decay during the aggregation process examined in this paper.

The large magnetic field also alters MQN interaction with each other through magnetic attraction and enhances their interaction with ordinary matter through the greatly-enhanced stopping power of the magnetopause around high-velocity MQNs moving through a plasma [17]. Searches [31] for quark nuggets with underground detectors would not be sensitive to highly magnetic ones. For example, the paper by Gorham and Rotter [24] about constraints on anti-quark nugget dark matter (which do not constrain quark-nuggets unless the ratio of anti-quark nuggets to quark nuggets is shown to be large) assumes that limits on the flux of magnetic monopoles from analysis by Price, *et al.* [32] of geologic mica buried under 3 km of rock are also applicable to quark nuggets. Gorham and Rotter also cite work by Porter, *et al.* [33-34] as constraining quark-nugget (nuclearite) contributions to dark matter by the absence of meteor-like objects that are fast enough to be quark nuggets. Bassan, *et al*. [35] looked for quark nuggets (nuclearites) with gravitational wave detectors and found signals much less than expected for the flux of dark matter. However, all of these experiments assumed the cross section for momentum transfer is the geometric cross section, which is many orders of magnitude smaller than the cross section of its magnetopause [17].